\documentclass[twocolumn]{article}

\usepackage[english]{babel}
\usepackage{booktabs}

\usepackage{authblk}
\usepackage[
    a4paper,
    top=2cm,
    bottom=2cm,
    textwidth=7.16in
]{geometry}

\usepackage{caption}
\usepackage{amsmath}
\usepackage{graphicx}
\usepackage[colorlinks=true, allcolors=black]{hyperref}
\usepackage{xcolor}
\usepackage{eurosym}
\usepackage{array}

\usepackage[capitalise]{cleveref}

\usepackage{titlesec}
\usepackage{ragged2e}

\titleformat{\section}
  {\normalfont\large\bfseries\RaggedRight}
  {\thesection}{0.5em}{}
\titlespacing*{\section}
  {0pt}{1.2em}{0.5em}

\titleformat{\subsection}
  {\normalfont\normalsize\bfseries\RaggedRight}
  {\thesubsection}{0.5em}{}
\titlespacing*{\subsection}
  {0pt}{1.0em}{0.4em}

\titleformat{\subsubsection}
  {\normalfont\normalsize\itshape\RaggedRight}
  {\thesubsubsection}{0.5em}{}
\titlespacing*{\subsubsection}
  {0pt}{0.8em}{0.3em}

\title{Bringing Chip Tapeout Into University Education}

\author{

Luca Pezzarossa$^{1}$,
Martin Schoeberl$^{1}$,
Matti K\"{a}yr\"{a}$^{2}$,
Nooshin Nosrati$^{3}$,\linebreak
Matthias Bo Stuart$^{1}$,
Timo D. H\"{a}m\"{a}l\"{a}inen$^{2}$,
Jean-Max Dutertre$^{4}$
and Michael Pehl$^{5}$
\vspace{1em}

{\normalsize
$^{1}$Department of Applied Mathematics and Computer Science,\linebreak
Technical University of Denmark, Kongens Lyngby, Denmark\\
\vspace{0.4em}
$^{2}$SoC Hub Research Centre,\linebreak
Tampere University, Tampere, Finland\\
\vspace{0.4em}
$^{3}$Department of Electronics \& Embedded Systems,\linebreak
KTH Royal Institute of Technology, Stockholm, Sweden\\
\vspace{0.4em}
$^{4}$Mines Saint-\'{E}tienne, CEA-Leti, Centre CMP, \linebreak \'{E}quipe Commune SAS, Gardanne, France\\
\vspace{0.4em}
$^{5}$TUM School of Computation, Information and Technology,\linebreak
Technical University of Munich, Munich, Germany\\
\vspace{1em}

\texttt{lpez@dtu.dk, masca@dtu.dk, matti.kayra@tuni.fi, nosrati@kth.se,}\\
\texttt{mbst@dtu.dk, timo.hamalainen@tuni.fi, dutertre@emse.fr, m.pehl@tum.de}
\vspace{0.7cm}
}
}

\date{}

\begin{document}
\maketitle

\begin{abstract}
Providing students with experience from chip specification to fabricated silicon can strengthen chip-design education, but integrating tapeout into regular teaching is difficult to scale, especially across universities with different curricula, schedules, regulations, and technical infrastructures.
This article discusses the implementation experience and lessons learned from a cross-university approach developed within the Edu4Chip European project.
The approach combines aligned learning outcomes, local Master's implementations, and a shared design-to-silicon framework across five European universities.
\end{abstract}

\vspace{2mm}

\section{Introduction}
\label{sec:introduction}

Increasing Europe's semiconductor capacity requires not only fabrication infrastructure but also engineers with the necessary skills.
A recent forecast estimates a shortfall of 65,000 skilled workers in Europe by 2030~\cite{ecsa_2025}.
In 2025, design engineers were among the hardest profiles to fill, while system architecture was the hardest-to-fill skill.
This shortage is attributed mainly to an aging workforce and limited growth in semiconductor-related graduates.
Policy recommendations therefore emphasize expanding Master's-level training, strengthening industry participation, and aligning university curricula with industry needs~\cite{ecsa_2025,glatter_2025}.

Chip design spans architecture, verification, physical implementation, fabrication, and testing.
University programs often provide practical experience through simulation and field-programmable gate array (FPGA) prototyping, but integrating application-specific integrated circuit (ASIC) fabrication into regular teaching is more demanding.
Tapeout adds cost, fixed deadlines, restricted tool and process access, and substantial coordination, particularly when several groups or universities contribute designs.

The challenge is therefore not simply to give students a one-time opportunity to design a chip, but to make design-to-silicon experience a repeatable part of university education.
Existing initiatives address parts of this problem through tapeout courses, fabrication services, open-source design flows, and multi-university coordination.
The scaling problem is how to combine these elements into a recurring model across institutions with different curricula, regulations, schedules, and technical infrastructures.

We address this problem through a federated model developed and implemented across five European universities within the Edu4Chip European project.
The model combines aligned learning outcomes with locally implemented Master's programs and specializations, a reusable chip platform, and common design infrastructure that lets student groups develop and integrate subsystems without building a complete system-on-chip from scratch.
Edu4Chip supports implementation through university courses, fabrication activities, mobility, and shared educational materials.

This article examines how learning goals can be aligned without identical curricula, how reusable infrastructure can support practical design-to-silicon activities, and what technical, administrative, and organizational challenges arise in practice.
Rather than comparatively evaluating learning outcomes, which requires a longer time horizon, we focus on the immediate question: \textit{What does it take to make design-to-silicon education feasible, repeatable, and scalable across heterogeneous university environments?}

\section{Real-Silicon Education: The Scaling Gap}
\label{sec:related_activities}

Strong educational programs in chip design already exist across electrical engineering, microelectronics, embedded systems, computer engineering, and integrated-circuit design.
Examples include established MSc or graduate programmes at EPFL\footnote{\url{https://www.epfl.ch/education/master/programs/electrical-and-electronic-engineering/}},
ETH Zurich\footnote{\url{https://ethz.ch/en/studies/master/degree-programmes/engineering-sciences/electrical-engineering-and-information-technology.html}},
TU Delft\footnote{\url{https://www.tudelft.nl/onderwijs/opleidingen/masters/ee/msc-electrical-engineering}},
Eindhoven University of Technology\footnote{\url{https://www.tue.nl/en/education/graduate-school/master-electrical-engineering}},
Politecnico di Milano\footnote{\url{https://www.polimi.it/en/education/laurea-magistrale-programmes/programme-detail/electronics-engineering}},
MIT\footnote{\url{https://www.eecs.mit.edu/}},
Stanford University\footnote{\url{https://ee.stanford.edu/}},
UC Berkeley\footnote{\url{https://eecs.berkeley.edu/}},
and TUM Asia in cooperation with Nanyang Technological University\footnote{\url{https://tum-asia.edu.sg/graduate-studies/master-integrated-circuit-design/}}.
A recent IEEE Spectrum article~\cite{mok2025semiconductor_courses} also describes rising student enrollment and new educational offerings in the United States.

Several initiatives go beyond conventional coursework by bringing students closer to real silicon. They broadly address three aspects of the problem: providing end-to-end tapeout experience, lowering barriers to fabrication and design infrastructure, and coordinating chip-design education across institutions.
Individual tapeout experiences demonstrate that students can be taken through substantial parts of the silicon-design flow. Examples include the Apple New Silicon Initiative at Georgia Tech\footnote{\url{https://ece.gatech.edu/apple-nsi}}
and the open-source SoC course experience around the Croc platform at ETH Zurich~\cite{zelioli2026croctraininggenerationchip}.
Croc is a particularly relevant comparison, where students work with a relatively compact, mature RISC-V platform and follow the design from RTL toward silicon.
Our setting instead integrates independently developed subsystems from different groups into a shared chip, adding requirements for common interfaces, verification, distributed development, and chip-level integration. 
In China, the ``One Student One Chip'' program\footnote{\url{https://ysyx.oscc.cc/en/project/intro.html}}
uses a massive open online course format in which participants learn to construct a RISC-V processor with both front- and back-end design steps.

Fabrication services and open infrastructure reduce cost, tooling, and access barriers. In Europe, Europractice\footnote{\url{https://www.europractice-ic.com/}} and CIME-P\footnote{\url{https://cime-p.cime.grenoble-inp.fr}}
provide access to design tools, training, and multi-project wafer services for universities. OpenLane\footnote{\url{https://openlane.readthedocs.io/en/latest/}}
showed how open-source tools such as Yosys, OpenROAD, Magic, Netgen, and KLayout can be connected into an automated RTL-to-GDSII flow. LibreLane\footnote{\url{https://github.com/librelane/librelane}}
continues this ecosystem as a more modular ASIC implementation flow, while Caravel\footnote{\url{https://github.com/chipfoundry/caravel}}
provides a reusable SoC harness for complete-chip integration. Tiny Tapeout\footnote{\url{https://tinytapeout.com/}}~\cite{tinytapeout2024}
further lowers the entry barrier by enabling beginners and students to fabricate small designs through low-cost shared tapeouts. SwissChips\footnote{\url{https://swisschips.ethz.ch/}}
builds on a similar idea by supporting Swiss residents in joining Tiny Tapeout runs and by offering workshops on open-source chip-design methods.
These efforts show that fabrication and implementation access, once a major barrier for educational tapeouts, can increasingly be treated as shared infrastructure rather than something every university must establish independently.

Multi-university initiatives address curriculum development, mobility, capacity, and links to industry. RESCHIP4EU\footnote{\url{https://28digital.eu/eu-collaborations/reschip4eu/}}
concentrates on establishing a chip-design-oriented double-degree MSc program. GreenChips-EDU\footnote{\url{https://www.tugraz.at/projekte/greenchips-edu/home}}
addresses sustainable microelectronics education through digital learning resources, student exchange, and MSc-level curriculum development. CHIPS of Europe\footnote{\url{https://chipsofeurope.eu/}}
similarly updates curricula, offers virtual laboratory environments, and promotes industry-relevant training. The European Chips Skills Academy\footnote{\url{https://chipsacademy.eu/}}
complements these activities through online learning, mobility, summer schools, micro-credentials, and links between education and industry.
RISC-V International maintains educational material for the RISC-V community\footnote{\url{https://riscv.org/community/training/}}
while the FOSSi Foundation\footnote{\url{https://fossi-foundation.org/}}
provides organizational support, events, and infrastructure for the open-source silicon community. Together, these initiatives also reflect the increasing European policy attention to semiconductor skills and the objectives of the European Chips Act~\cite{eu_chips_act_web}.

These developments address important parts of the problem: complete silicon flows, more accessible fabrication, and coordinated curricula and resources.
Scaling real-silicon education across universities, however, requires these elements to work together across different curricula and regulations, supported by reusable infrastructure, access to tools and fabrication, distributed development, chip-level integration, and post-silicon testing.
As also argued by Krupp et al.~\cite{krupp_edu}, broader adoption requires institutional support and coordinated access to tools, technologies, fabrication, and testing.
The remaining challenge is therefore how to combine these ingredients into a repeatable model across heterogeneous universities.

\section{A federated model for scaling chip-design education}
\label{sec:program_design}

We use a federated model that aligns common learning outcomes and practical design-to-silicon experience without enforcing an identical curriculum across participating universities.

\subsection{Edu4Chip in brief}

The model is being implemented through Edu4Chip, a four-year European collaboration involving the Technical University of Munich (TUM), KTH Royal Institute of Technology, the Technical University of Denmark (DTU), Tampere University (TAU), and the École des Mines de Saint-Étienne of Institut Mines-Télécom (IMT), together with research and industry partners (Fraunhofer IIS, LogiqWorks, MINRES, and SyoSil).
The main objective of the project is to improve existing Master's programs in chip design and microelectronics and to start new programs.
The university partners coordinate Master's-level chip-design education and practical design-to-silicon activities, while concrete course structures, regulations, and technical emphases remain local.

For scalability, we follow a competence-centric rather than a one-size-fits-all approach.
The universities define common base skills, especially the capability to develop a chip in practice.
Students can take a design from specification to tapeout and, where possible, test it after manufacturing, while universities retain their individual strengths and accommodate local legal and administrative requirements.
Scaling this approach also requires suitable technical infrastructure.

\subsection{Common outcomes, local implementations}

The local Master's implementations are constructed around a chip-design tapeout project and target common learning competences while allowing each university to organize its program and specializations according to local needs.

To prepare students for a career in chip design, certain qualifications are considered mandatory.
These include an advanced understanding of the physical relationships and functionality of integrated circuits, the capabilities and underlying methods of design tools, and the capability to apply this knowledge to design innovative and efficient integrated circuits.
These competences are common to specializations ranging from analog/mixed-signal and digital design to system design and verification.
Since students may enter with different Bachelor's-level backgrounds, the core competences are addressed early, including training in hardware description languages or analog design tools.

Students can then specialize in analog/mixed-signal design, digital design for security, artificial intelligence, or asynchronous circuits, system and hardware/software co-design, and test and verification.
These areas reflect industry needs and university research strengths, supported by industry experts and guest lecturers.

The practical path starts with chip-design tools and description languages, and leads to a tapeout project covering specification, functional and physical design, and tapeout.
Students may implement an application-specific block and integrate it into the shared SoC under defined interface, area, frequency, and optimization constraints.
They develop verification strategies, make hardware/software partitioning decisions, may validate the design on an FPGA, and after manufacturing perform bring-up and evaluation.
The project therefore connects digital design, computer architecture, verification, physical implementation, and embedded software around one artifact.

\begin{figure}[!t]
    \centering
    \includegraphics[width=0.95\columnwidth]{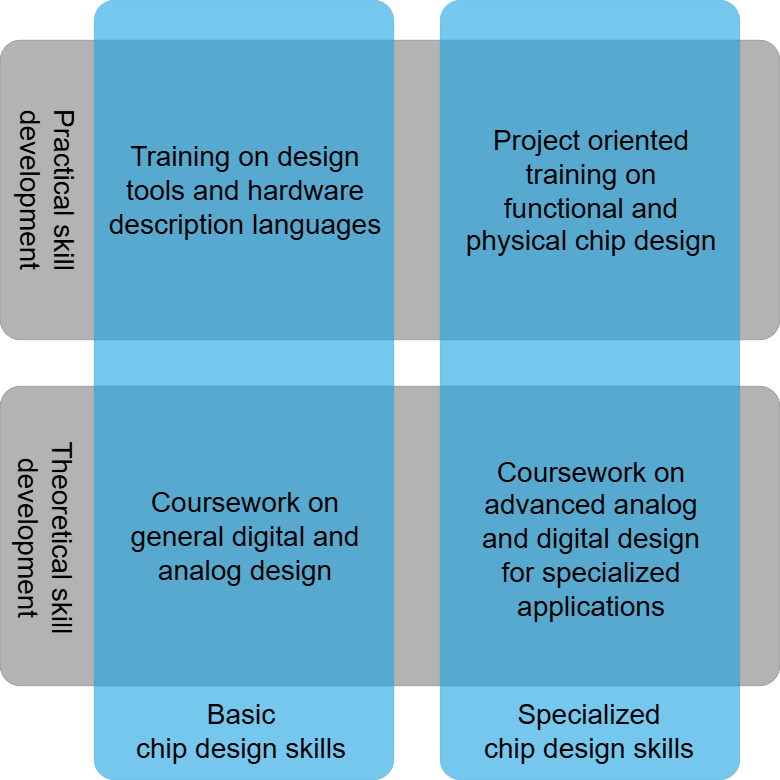}
    \caption{Overview of the common Master's-level educational framework for technical skill development. Theoretical and practical training progress from basic to specialized topics, with practical activities advancing from training on design tools and hardware description languages to project-oriented functional and physical chip design.}
    \label{fig:edu4chip_framework}
\end{figure}

The framework can be seen as a matrix structure as shown in Figure~\ref{fig:edu4chip_framework}: from left to right, students progress from basic to specialized chip-design skills, while theoretical and practical skill development complement each other.
Teaching modules can be placed within this structure according to local requirements, giving a high degree of flexibility to the overall implementation.
Table~\ref{tab:local_programs} summarizes representative implementations at the five partner universities and illustrates how the common framework is adapted to different local curricula and technical strengths.
The current implementations realize the chip-design project as a single course from specification to tapeout, as a sequence of laboratories for functional and physical design, or as laboratories integrated into other modules, while maintaining the same overall learning goal.

This flexibility is also important for tool and technology access.
Commercial design flows may impose non-disclosure, licensing, or other legal restrictions that prevent access for some students or institutions.
The participating universities therefore use a combination of open-source tools and technologies, and commercial tools and advanced process nodes.

\begin{table*}[t]
\centering
\small
\caption{\label{tab:local_programs}Representative local implementations of the framework, illustrating different chip-design paths and complementary technical strengths across the partner universities.}
\vspace{0.0em}
\begin{tabular}{@{}
>{\raggedright\arraybackslash}p{0.1\textwidth}
>{\raggedright\arraybackslash}p{0.415\textwidth}
>{\raggedright\arraybackslash}p{0.415\textwidth}
@{}}
\toprule
\textbf{University} & \textbf{Representative chip-design path} & \textbf{Representative technical strengths} \\
\midrule
TUM &
Fundamental lab \!$\rightarrow$\! functional-design research lab \!$\rightarrow$\! physical-design research lab \!$\rightarrow$\! test and evaluation &
Analog/mixed-signal and digital design; physical design; security and application-specific circuits \\
\midrule
KTH &
Digital design and verification \!$\rightarrow$\! ASIC/FPGA design \!$\rightarrow$\! project course in application-specific ICs &
ASIC/FPGA design; analog and RF ICs; embedded systems and verification \\
\midrule
DTU &
Digital systems and introductory chip design \!$\rightarrow$\! VLSI, verification, and test \!$\rightarrow$\! integrated chip-project activities &
Digital systems; computer architecture; hardware/software co-design; verification and open-source flows \\
\midrule
TAU &
Digital design \!$\rightarrow$\! Logic synthesis \!$\rightarrow$\! System-on-chip design and verification \!$\rightarrow$\! Chip implementation &
SoC design; verification; high-level synthesis; backend implementation \\
\midrule
IMT &
Electronics, HDL, and architecture \!$\rightarrow$\! FPGA/ASIC projects \!$\rightarrow$\! microelectronics design and tapeout &
Semiconductor devices and manufacturing; analog/digital/secure IC design; FPGA/ASIC design \\
\bottomrule
\end{tabular}
\end{table*}

\subsection{Learning through a shared silicon project}

The design-to-silicon activity is the practical core of the model, where theory, laboratory training, and teamwork come together. Students work on a subsystem that becomes part of a manufactured chip. 
Because each subsystem forms part of a larger chip, teams have to specify and document their designs and deliver them in time for integration.
Students therefore encounter the organizational side of engineering, including planning, review, coordination, and dependence on the work of others.

Courses in digital design, computer architecture, analog circuits, and verification give students the concepts they need, while laboratory exercises train individual steps of the design flow, usually one tool or method at a time. The chip project supplies the wider context: the same skills are used in sequence toward a physical result, and students can see how decisions made early in the work affect later stages. 
The shared silicon project thus provides a common practical experience across otherwise different local implementations.

\section{Technical foundations for scalable design-to-silicon education}
\label{sec:tapeout}

Implementing a chip exposes students to the complete engineering cycle from specification and verification to physical implementation, fabrication, and testing.
The same complexity makes this difficult to accommodate within a course, especially when several groups or universities contribute designs.
We address it through a reusable chip platform, standardized integration mechanisms, and an educational flow from specification to fabricated silicon.

\subsection{A reusable platform for distributed development}

\begin{figure}[!t]
    \centering
    \includegraphics[width=0.98\columnwidth,trim={2.8mm 0 0 0},clip]{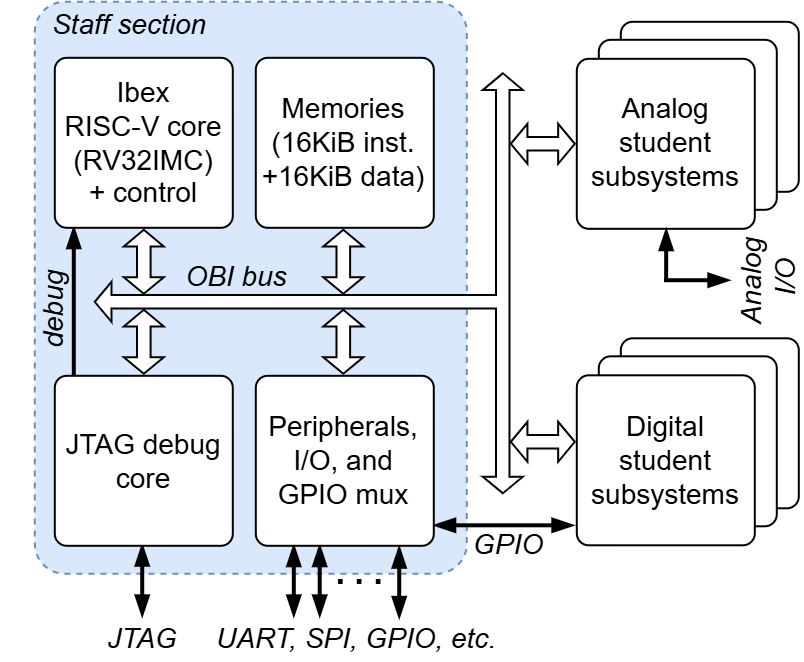}
    \caption{Simplified architecture of the Didactic SoC platform. The reusable staff section contains the Ibex RISC-V core and control logic, instruction and data memories, JTAG debug support, and peripheral and I/O infrastructure. Independently developed digital and analog student subsystems connect through the internal interconnect and shared I/O interfaces.}
    \label{fig:didactic_soc}
\end{figure}

To support practical design-to-silicon education across institutions, we developed a reusable open-source baseline chip platform, called Didactic SoC\footnote{\url{https://github.com/Edu4Chip/Didactic-SoC}}.
The platform separates a fixed staff-designed section from independently developed student subsystems.
The staff section provides the common infrastructure required for a complete system, including processor-based software execution, memories, debug access, peripherals, clock and reset distribution, and standardized interfaces to the student subsystems.
This allows student groups to concentrate on well-defined designs without requiring each group or university to build a complete SoC from scratch, while the common infrastructure remains operational independent of the maturity of individual subsystems. Figure~\ref{fig:didactic_soc} shows the overall organization of the Didactic SoC.

Because the universities use different toolchains, teaching methods, and local infrastructures, interoperability must be addressed explicitly.
Subsystem interfaces are described using IP-XACT and integrated with the Kactus2 tool~\cite{Kamppi2017Kactus2}.
Each student subsystem follows a common wrapper template and programmable interface, allowing digital and mixed-signal subsystems to be developed independently with local tool setups while preserving compatibility with the top-level design.
Separate reset and clock-gating controls isolate the subsystems so that faults in one design do not prevent the rest of the chip from being tested.
Backup designs are also provided for subsystems that do not meet integration or manufacturing requirements before the tapeout deadline.
This isolation is important in an educational setting, where subsystem maturity can vary substantially between student groups and a failure in one design should not invalidate the learning outcome or testability of all others.

Design artifacts are exchanged through a Git-based hierarchical repository.
Git is used for version control and Bender for dependency management, while build automation uses Makefiles together with Tcl, Python, and shell scripts.
IP-XACT provides a common machine-readable format for interfaces, connectivity, and memory maps, and is used through Kactus2 to create system-level integration and structural RTL generation.
Together with versioned dependencies and common subsystem wrappers, this provides a reproducible integration environment while allowing local development to remain flexible.

The framework supports both commercial and open-source toolchains.
Partners have used Siemens EDA Questa for simulation, commercial back-end flows, and open-source tools including Verilator, PyUVM, Yosys, LibreLane, Magic, and KLayout, together with the SkyWater 130 PDK.
Supporting both environments improves accessibility while preserving exposure to industrial design flows.

\subsection{From specification to fabricated silicon}

The educational flow follows the main phases of industrial chip development.
Students are first introduced to scripting and automation, hardware description and verification, version control, build systems, and simulation through small exercises that prepare them for the target chip.

The front-end phase begins with specifications defining functionality, requirements, and external interfaces, followed by architectural models and register-transfer level (RTL) implementations using SystemVerilog and reusable intellectual property (IP).
Verification is performed at subsystem and SoC levels: reusable environments test student subsystems, while top-level simulation tests interfaces, processor software, and interactions with integrated subsystems. JTAG supports debug and control.
FPGA prototyping adds pre-silicon validation of boot, hardware-software interaction, and physical peripherals, showing that successful subsystem simulation is necessary but not sufficient for a working chip.

After functional verification, the designs are transformed into a physical layout.
This phase includes logic synthesis, floorplanning, placement, routing, timing closure, and physical verification.
Students learn how the semiconductor technology described by the process design kit (PDK), including standard-cell libraries, I/O libraries, memory macros, and design rules, constrains the implementation.
Sign-off includes design-rule checking (DRC), layout-versus-schematic checking (LVS), timing analysis, and equivalence checking.
Design-for-test concepts are introduced to show how manufacturability and production testing are incorporated into chip development.

Fabrication is treated as part of the complete design-to-silicon sequence rather than as its endpoint.
We use multi-project wafer (MPW) runs for cost-effective fabrication.
The manufactured chips can be packaged and mounted on dedicated printed circuit boards (PCBs), while bare dies can also be bonded directly to a carrier PCB using chip-on-board assembly.

Post-silicon validation begins with power, clock, and reset checks, followed by JTAG and software-based subsystem testing.
The Didactic SoC provides debug, software execution, UART, SPI, and GPIO interfaces for bring-up and characterization, allowing students to compare measurements with pre-silicon predictions.
Treating bring-up as part of the educational flow means that test firmware, board interfaces, packaging, and external-signal access must be considered before tapeout.
A jointly developed PCB supports this validation.

\section{First implementations and outcomes}
\label{sec:outcomes}

The first implementations show the model moving from planning to practice.
The common platform has supported tapeout activities, while open-source runs demonstrated a lower-cost route to fabrication.
Student designs have reached tapeout and fabrication submission, and summer schools and workshops have provided additional entry points.

\subsection{Platform validation and first tapeouts}

The first joint tapeout within Edu4Chip used the Didactic SoC baseline in a commercial GF 22~nm process with commercial EDA tools.
In this first iteration, the subsystem slots were populated by four digital and one analog partner-developed designs, providing a validation of the platform and integration flow before student-created subsystems enter later shared tapeouts.

Separately, DTU targeted a lower-cost open-source path in the SkyWater 130~nm process using an open-source design flow and the chipIgnite template.
A first test chip was used to validate the flow, followed in Spring 2026 by a student chip developed in the course ``Introduction to Chip Design''.
For the DTU run, the fabrication slot cost approximately \euro15,000.

KTH ran a chip-design course in which a student design was taken from specification and RTL through verification and physical implementation to sign-off and submission in GF 22~nm using commercial EDA tools.

Finally, in summer 2025, IMT taped out a student-designed fully custom digital ASIC in X-FAB's 180~nm technology using commercial EDA tools. The fabricated chip was electrically validated, exercising the complete design-to-silicon process. 

Together, these implementations used both advanced and legacy commercial processes as well as a lower-cost open-source route.
They show that the same educational objective can be reached through different process and tool choices, from simulation and integration to student-created tapeouts.

\subsection{A practical implementation at DTU}

After the test tapeout, DTU ran the course ``Introduction to Chip Design'' in Spring 2026, centered on a shared student tapeout.
Twenty students worked in six groups and contributed to a common repository.
The students had completed two semesters of digital electronics, a digital-design project, computer architecture, and an FPGA project implementing a pipelined RISC-V processor.
Initial laboratory exercises introduced LibreLane and a Caravel design containing a simple Wishbone-connected circuit.
The students then added an I/O port and integrated the Wildcat~\cite{wildcat:arcs2025} RISC-V processor.
After these exercises, each group selected a subproject.
Assessment was based on contributions to the shared design, participation in problem solving, and a final presentation.

\begin{figure}[!t]
  \centering
  \includegraphics[width=0.98\columnwidth]{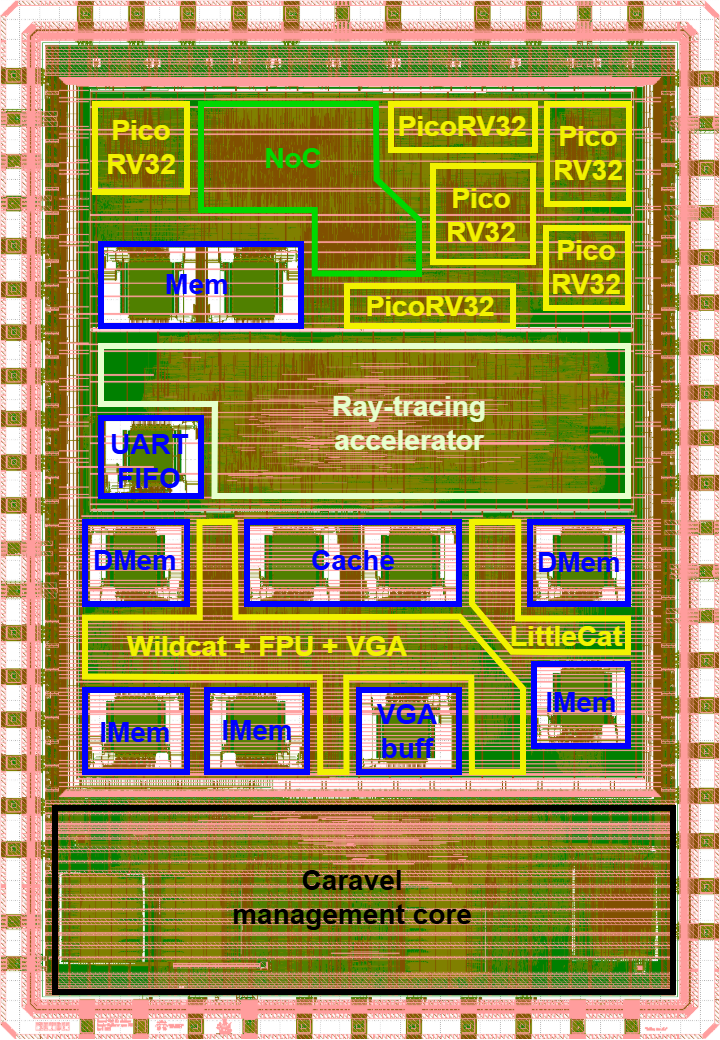}
  \caption{Annotated physical layout of the $\mathrm{10\;mm^2}$ DTU student chip in SkyWater 130~nm. The upper region contains the six-core PicoRV32/S4NoC subsystem, while the central and lower regions contain the Wildcat-based subsystem, including its memories, cache, VGA controller, and ray-tracing accelerator. The Caravel management core occupies the bottom of the die.}
  \label{fig:student:chip}
\end{figure}

The submitted $\mathrm{10\;mm^2}$ design targets the SkyWater 130~nm process and contains two main subsystems and 11 OpenRAM memories.
The first is based on Wildcat and includes instruction and data memories, a data cache, VGA terminal functionality, a ray-tracing accelerator, SPI, and a floating-point unit, with its peripherals and accelerators memory-mapped and connected to Caravel through Wishbone.
A smaller CPU, named LittleCat, is also present.
The second subsystem contains six PicoRV32 cores\footnote{\url{https://github.com/YosysHQ/picorv32}}, two scratchpad memories, and the S4NoC network-on-chip~\cite{t-crest:s4noc, s4noc:nocarc2019}, accessed through a UART-based boot interface; an additional UART-booted Wildcat core is also included.
The chip physical organization is shown in Figure~\ref{fig:student:chip}.

GitHub Actions were used for RTL testing and full-chip hardening.
Hardening was performed hierarchically: first selected accelerators and subsystems, then the main student design containing these macros, and finally the project wrapper.
Separately hardened blocks therefore had to retain stable interfaces before they could be incorporated into the next level of the hierarchy, mirroring staged design freezes in larger SoC projects.
The resulting GDSII passed DRC, LVS, and timing checks and was submitted to ChipFoundry for fabrication in May 2026.

The final layout reflects the distributed course structure: independently developed components had to be stabilized, hardened, and assembled under one tapeout deadline.
The reusable framework and open flow enabled several groups to contribute to one fabricated design, while repository organization, stable interfaces, continuous integration, and instructor involvement became important as subsystems converged.
An informal course survey indicated strong interest but also that the project could feel complex initially; comments included ``The course is fun and cool'' and ``The coolest course I've had so far.''

\subsection{Summer schools and different entry points}

We organized two summer schools providing short-format entry points into chip design through lectures, academic and industry keynotes, and hands-on activities.
At the first summer school, hosted by DTU in August 2025, 45 participants from eight European universities followed the chip-design flow from specification to testing and were introduced to open-source tools.
At the second, hosted by IMT in August 2026, 51 participants from seven universities were introduced to the fundamentals of chip design and the full custom design of standard cells. They successfully designed several IP blocks, including an SRAM cell, a ring oscillator, and a SPI communication block interfacing an ALU.
Their IPs are being embedded in an X-FAB 180 nm ASIC to be sent for manufacturing.
Together, the two editions gave students from different backgrounds a compact introduction to the design-to-silicon process used in the degree programs.

Additional outreach included Tiny Tapeout workshops, which attracted more than 80 participants in 2025 and 72 in 2026, as well as fairs and information events for prospective students.
Together, these activities broaden participation and provide routes into more substantial chip-design education.

\section{Lessons learned for scaling real-silicon education}
\label{sec:conclusion}

Making design-to-silicon education repeatable across universities requires curriculum design, technical infrastructure, tool access, project organization, and post-silicon activities to be considered together.
To conclude, we present the following lessons, which summarize what has worked in practice and where challenges remain.

\textbf{Align outcomes, not curricula.}
The partner universities differ in program structure, ECTS allocation, prerequisites, schedules, and technical specialization.
Aligning learning objectives and practical competencies proved more realistic than creating identical curricula.
Cross-university teaching also has to accommodate grade recognition, examination rules, access to learning platforms, data protection, and different academic calendars.
For example, for a given shared MPW run, the fixed submission deadline falls at different points in the partners' teaching periods, making a single course schedule impractical.
Instead, we align on common deliverables and design-freeze dates while each university organizes the preparatory work within its local calendar.

\textbf{Use a reusable chip platform to manage complexity.}
A common platform allows students to develop well-defined subsystems without requiring every group to build a complete SoC, while still engaging with realistic requirements for interfaces, verification, physical implementation, and chip-level integration.
Reusable interfaces, wrappers, and verification infrastructure reduce the effort required to integrate independently developed designs and make the approach transferable between universities.
Reusability still requires maintenance: repositories, reference designs, verification infrastructure, tool versions, dependencies, documentation, and CI infrastructure must be maintained between offerings.

\textbf{Shared tapeouts need explicit integration ownership.}
Our experience highlighted the importance of a dedicated integration and tapeout coordination role.
Maintaining the top-level design and repository, tracking verification and dependencies, managing the design freeze, and coordinating the MPW submission require centralized ownership in addition to subsystem contributors.
This role closely resembles system-level integration in industrial SoC projects and we found it best assigned to an experienced doctoral researcher, postdoctoral researcher, or staff member.

\textbf{Tool access must be considered from the start.}
Commercial EDA tools and process technologies expose students to industrial practice, but licenses, NDAs, PDK restrictions, server access, and export controls can limit participation and require institutional support.
Open-source flows provide an alternative: students can often use the complete flow on their own machines, while designs, scripts, PDK interfaces, and teaching materials are easier to share.
For example, the DTU ``Introduction to Chip Design'' material is public\footnote{\url{https://github.com/os-chip-design/chip-design-intro}}, and we have started an open-source textbook on chip design~\cite{chip:design:book}.

\textbf{Plan beyond tapeout.}
Packaging, PCB design, firmware, bring-up, testing, and characterization require additional infrastructure and time and should be planned together with the chip project rather than after fabrication.
For example, the Didactic SoC uses low-voltage I/Os, requiring level shifting on the test PCB to interface with standard 3.3-V peripherals; such board-level requirements need to be identified early to avoid delaying post-silicon bring-up.

For institutions considering a similar approach, scalable real-silicon education requires more than fabrication access. 
Adaptable curricula, reusable technical scaffolding, accessible tool paths, clear integration ownership, post-silicon planning, and continued maintenance are essential to make individual tapeouts into a repeatable educational activity.

\section*{Acknowledgments}
This work was funded by the European Union through the European Health and Digital Executive Agency (HaDEA) under Grant Agreement No.~101123086 (Edu4Chip). Views and opinions expressed are, however, those of the authors only and do not necessarily reflect those of the European Union or HaDEA. Neither the European Union nor the granting authority can be held responsible for them.

OpenAI ChatGPT (GPT-5.6 Sol) was used as an editorial aid during manuscript preparation, including for restructuring the manuscript, language refinement, condensation, and rephrasing.
All technical content, claims, references, and final wording were reviewed and approved by the authors, who take full responsibility for the manuscript.

\section*{Biographies}

\noindent\textbf{Luca Pezzarossa} is an associate professor at DTU Compute, Technical University of Denmark. His research interests include reconfigurable systems, computer architecture, and FPGA acceleration. He received his PhD from DTU in 2017 and is an IEEE member.

\vspace{0.5em}
\noindent\textbf{Martin Schoeberl} is a professor at DTU Compute, Technical University of Denmark. His research interests include time-predictable computer architecture, real-time systems, and safety-critical systems. He received his PhD from Vienna University of Technology in 2005 and is an IEEE member.

\vfill
\pagebreak

\noindent\textbf{Matti K\"{a}yr\"{a}} is a doctoral researcher at Tampere University's SoC Hub Research Centre. His research interests include SoC design, RISC-V multiprocessors, and ASIC implementation. He received his master's degree from Tampere University in 2019. 

\vspace{0.5em}
\noindent\textbf{Nooshin Nosrati} is a postdoctoral researcher at KTH Royal Institute of Technology. Her research interests include hardware design and modeling, reliability, and digital-system testability. She received her PhD from the University of Tehran in 2024 and is an IEEE member.

\vspace{0.5em}
\noindent\textbf{Matthias Bo Stuart} is an associate professor at DTU Compute, Technical University of Denmark. His research interests include computer architecture and system engineering. He received his PhD from DTU in 2010.

\vspace{0.5em}
\noindent\textbf{Jean-Max Dutertre} is a professor at the Centre Micro\'{e}lectronique de Provence, Mines Saint-\'{E}tienne. His research interests include hardware security, fault-injection attacks, and countermeasures. He received his PhD from the University of Science of Montpellier in 2002 and is an IEEE member.

\vspace{0.5em}
\noindent\textbf{Timo D. H\"{a}m\"{a}l\"{a}inen} is a professor and head of the SoC Hub Research Centre at Tampere University. His research interests include SoC design, design automation, and RISC-V and edge architectures. He received his PhD in 1997 and is an IEEE member.

\vspace{0.5em}
\noindent\textbf{Michael Pehl} is an adjunct teaching professor at the Technical University of Munich. His research interests include physical unclonable functions, side-channel and fault-injection analysis, and secure hardware. He received his Dr.-Ing. from TUM in 2012 and is an IEEE member. 

\vfill

\end{document}